\documentclass[3p, 11pt]{elsarticle}
\usepackage{graphicx,amscd,amsmath,amssymb,verbatim}
\usepackage{amsfonts,epsfig}
\usepackage{mathptmx}
\usepackage{setspace}
\usepackage{epsfig}
\usepackage{url}
\usepackage{multirow}
\usepackage{color}

\begin{document}

\journal{Elsevier}

\begin{frontmatter}

\title{Waterlike Features, Liquid-Crystal Phase and Self-Assembly in Janus Dumbbells}

\author{Jos\'e Rafael Bordin$^1$} 

\address{1-Campus Ca\c capava do Sul, Universidade Federal
do Pampa, Av. Pedro Anuncia\c c\~ao, 111, CEP 96570-000, 
Ca\c capava do Sul, RS, Brazil

{\normalsize{E-mail:josebordin@unipampa.edu.br}}}

\begin{abstract}

We use Molecular Dynamics simulations to explore the properties
of dimeric Janus nanoparticles. The nanoparticles are modeled
as dumbbells. One kind of monomer interacts by a Lennard Jones potential,
while the other specie of monomer interacts though a two length scale potential.
This specific two length scale potential do not present waterlike
anomalies in bulk. However, our results shows that the combination
of teo length scales potential and LJ potential in
the Janus nanoparticle will lead to thermodyncamic and dynamic anomalies.
The self-assembly properties were also explored. We observe distinct
kinds of self-assembled structures and a liquid-crystal phase. This results indicates
that is possible to create Janus nanoparticles with waterlike features
using monomers without anomalous behavior. The anomalies and structures
are explained with the two length scale potential characteristics.

\end{abstract}

\end{frontmatter}

\setlength{\baselineskip}{0.7cm}

\section{Introduction}

\label{Section:Introduction} 
Most part of the materials contracts on cooling at constant pressure. 
Also, we expect that the diffusion coefficient will decrease when
the pressure, or density, of the system is increased. However, this is not the case from the so called
anomalous fluids. The most well known anomalous fluid is water~\cite{Ke75,An76}, 
with 73 known anomalies~\cite{URL}. Liquid water at 1.0 atm have a maximum in the density at 
temperature of $4^o$C, expanding as is cooled to $0^o$C.
Others materials, as
silicon~\cite{Sa03}, silica~\cite{Sh06},
Te~\cite{Th76}, Bi~\cite{Handbook}, 
Si~\cite{Ke83}, $Ge_{15}Te_{85}$~\cite{Ts91},  liquid 
metals~\cite{Cu81}, graphite~\cite{To97} and 
$BeF_2$~\cite{An00}
also presents thermodynamic anomalies.
Water~\cite{Ne02a}, silicon~\cite{Mo05} and silica~\cite{Sa03} 
show a anomalous diffusion, characterized by a
maximum in the diffusion coefficient at constant temperature.

Core-softened (CS) potentials with two length scales (TLS)
have been widely used to study the general properties and
characteristics of fluids with anomalous behavior
~\cite{Jagla99, Oliveira08, Silva10, Fomin11}.
TLS potentials are characterized by having two preferred particle-particle
separations, while one length scale (OLS) potentials, as the Lennard Jones (LJ) potential,
shows only one preferred separation.

TLS potentials are able to reproduce waterlike anomalies in qualitative way if exists competition
between the two characteristic distances~\cite{Silva10, Barraz09, Oliveira10}.
However, if the energy penalty to the particle move from one scale to another is higher than the 
particle kinetic energy, the particle will get trapped in one length scale, and there will be no
competition and, as consequence, there will be no anomalous behavior. 

Another system of interest are colloidal suspensions. Despite the absence of 
anomalous behavior, TLS potentials were obtained from experiments using inversion of structural data
of colloidal systems~\cite{QP01, CA10}. Since such systems do not show waterlike anomalies,
they can be studied using a TLS potential without competition between the scales.

Particularly, colloidal Janus nanoparticles have
attract the attention of scientists
due the large range of applications of this new materials,
as applications in medicine, catalysis, photonic crystals,
stable emulsions, biomolecules and self-healing materials~\cite{Talapin10,ElL11, TuP13, WaM08,
WaM13, Zhang15}.
Dumbbells colloids are formed by two spheres
that overlap with a separation that varies from an almost total overlap
to one or two monomer diameters. The molecule anisotropy
plays quite a relevant role. The  properties of the system depend 
on the interaction potential that varies with their spatial separation
and  their relative orientations. 
In the case of Janus dumbbells~\cite{Yin01, SiC14, Lu02, YoL12},
each monomer have distinct characteristics, as charged/neutral or hydrophilic/hydrophobic. 
The competition between attractive and repulsive forces lead to the 
formation of self-assembly lamellae or
micellae phases~\cite{Li12}.

Recently, the production of silver-silicon (Ag-Si)~\cite{SiC14},
silica-polystyrene (SiO$_2$-PS)~\cite{Liu09} and
tantalum silicide-silicon (TaSi$_2$/Si)~\cite{Nomoeva15}
hybrid Janus dimers were reported.
Silicon and silica are classified as anomalous fluids,
and the silicon-silicon or silica-silica interaction in 
the pure system can be modeled by a TLS potential with competition
between the scales. The others monomers
can be described by an one length scale potential and consequently 
does not show the presence of the waterlike thermodynamic and dynamic
anomalies. In our previous work we have show that, if the TLS potential shows
dynamical and thermodynamical anomalous behavior for the monomeric case, 
the Janus dimer will have anomalies accordingly with the non-anomalous monomer properties~\cite{BoK15c}.

A new question arises when none of the monomers are anomalous, but one can be modeled with a TLS potential
without competition between the scales.
For instance, soft colloidal~\cite{Deng15, Ku15}, metallic/polymer~\cite{Jia12, Hu12}
and liquid-crystal/polymer~\cite{Jeong15} Janus dumbbells have a colloidal monomer whose
interaction can be described by a TLS potential~\cite{Vilaseca11} 
and a monomer that have OLS interaction.

In order to answer
this question  we explore the pressure
versus temperature phase 
diagram of a model systems. The system is a composed
by Janus particles in which one monomer interacts
through a TLS potential without anomalies
and the other monomer interacts
through a LJ potential. We investigate 
how the presence of the LJ monomer
affects the competition between the two characteristic distances,
the self-assembly structures and describe qualitatively
the thermodynamic phases.

The paper is organized as follows. The model, the methods 
and simulation details are described in the Section~\ref{Model}; the results and 
discussion are given in the Section~\ref{Results}; and 
then the conclusions are presented in the Section~\ref{Conclu}.

\section{The Model and the Simulation details}
\label{Model}
In this paper all the physical quantities are computed
in the standard LJ units~\cite{AllenTild},
\begin{equation}
\label{red1}
r^*\equiv \frac{r}{\sigma}\;,\quad \rho^{*}\equiv \rho \sigma^{3}\;, \quad 
\mbox{and}\quad t^* \equiv t\left(\frac{\epsilon}{m\sigma^2}\right)^{1/2}\;,
\end{equation}
for distance, density of particles and time , respectively, and
\begin{equation}
\label{rad2}
p^*\equiv \frac{p \sigma^{3}}{\epsilon} \quad \mbox{and}\quad 
T^{*}\equiv \frac{k_{B}T}{\epsilon}
\end{equation}
for the pressure and temperature, respectively, where $\sigma$ is the 
distance
parameter, $\epsilon$ the energy parameter and $m$ the mass parameter.
Since in the present work all physical quantities are expressed in
reduced LJ units we will omit the $*$ for simplicity.

\begin{figure}[ht]
 \begin{center}
 \includegraphics[width=8cm]{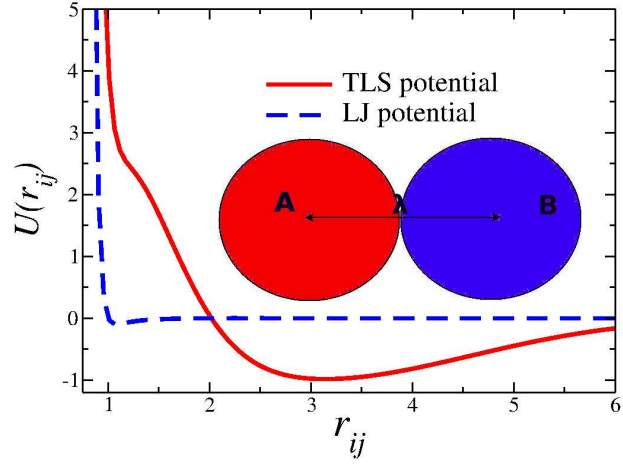}
 \end{center}
 \caption{Interaction potentials used in our simulations: the 
TLS potential (solid red line) and the LJ potential (dashed
blue line). Inset: Janus nanoparticles formed by A-B monomers.}
 \label{fig1}
 \end{figure}

$N$ dimers were used in each simulation, in a total of $2N$ particles.
Each monomer have diameter $\sigma$ and mass $m$, and the
Janus dumbbells are constructed using two monomers linked
rigidly at a distance $\lambda = \sigma$. The only difference
between the monomers are their interaction. Monomers
of type A interacts with another A monomer through a 
TLS core-softened potential,~\cite{Silva10}
\begin{equation}
\frac{U^{{\rm TLS}}(r_{ij})}{\varepsilon} = 4\left[ \left(\frac{\sigma}{r_{ij}}\right)^{12} -
\left(\frac{\sigma}{r_{ij}}\right)^6 \right]
+ u_0 {\rm{exp}}\left[-\frac{1}{c_0^2}\left(\frac{r_{ij}-r_0}{\sigma}\right)^2\right]- u_1 {\rm{exp}}\left[-\frac{1}{c_1^2}\left(\frac{r_{ij}-r_1}{\sigma}\right)^2\right]\;,
\label{AlanEq}
\end{equation}
where $r_{ij} = |\vec r_i - \vec r_j|$ is the distance between two A 
particles $i$ and $j$.
This equation has three terms: the first one is the standard LJ 12-6 potential~\cite{AllenTild}, the second is a Gaussian
centered at $r_0 = 0.70$, with depth $u_0 = 5.0$ and width $c_0 = 1.0$,
response for the shoulder in the potential shape,
and the last term is also a Gaussian, but centered at $r_1 = 3.0$, with depth 
$u_1 = -1.0 $ and width $c_1 = 0.5$, responsible for the attractive tail of the potential,
as indicated in figure~\ref{fig1} by the solid red line. Despite the presence of the 
two length scales, monomeric systems interacting through this potential did not 
shows waterlike anomalies since there was not observed competition
between the scales~\cite{Silva10}. B monomers interacts though a LJ potential,
 \begin{equation}
 \label{LJ}
 U^{\rm{LJ}}(r_{ij}) = \frac{4}{3}\cdot2^{2/3}\left[ \left(\frac{\sigma}{r_{ij}}\right)^{24} -
\left(\frac{\sigma}{r_{ij}}\right)^6 \right]\;,
 \end{equation}
 \noindent the blue dashed line in figure~\ref{fig1}. This particular shape for the LJ
 potential was chosen due to his small attractive well. Interactions
 between A and B monomers are also given by equation~\ref{LJ}.

$NVT$ Molecular dynamics simulations using the ESPResSo package~\cite{espresso1, espresso2}
were performed in order to obtain the pressure 
versus temperature ($p\times T$) phase diagram.
A total of $N=1200$ particles were used in our simulations. To ensure our results, simulations with
up to 9000 dimers where carried out, and essentially the same results were
obtained. The number density is defined as $\rho = N/V$, where $V=L^3$ is 
the volume of the cubic simulation box.
Standard periodic boundary conditions are applied in all directions. 
The system temperature was fixed using the Langevin thermostat 
with $\gamma = 1.0$. 
The equations of motion for the fluid particles were integrated
using the velocity Verlet algorithm, with a time step $\delta t = 0.01$.
The cutoff radius for the TLS potential is $r_{\rm cutTLS} = 6.5$, and
$r_{\rm cutLJ} = 2.5$ for the LJ potential.
We performed $5\times10^6$ steps to equilibrate the system. 
These steps are then followed by $5\times10^6$ steps for the results 
production stage. To ensure that the system was equilibrated, the pressure, kinetic energy
and potential energy were analyzed as function of time, as well several 
snapshots at distinct simulation times.
    
To study the dynamic anomaly the relation between 
the mean square displacement (MSD) with time is analyzed,  namely
\begin{equation}
\label{r2}
\langle [\vec r_{\rm cm}(t) - \vec r_{\rm cm}(t_0)]^2 \rangle =\langle 
\Delta \vec r_{\rm cm}(t)^2 \rangle\;,
\end{equation}
where $\vec r_{\rm cm}(t_0) = (x_{\rm cm}(t_0)^2 + y_{\rm cm}(t_0)^2 
+ z_{\rm cm}(t_0)^2)^{1/2} $ 
and  $\vec r_{\rm cm}(t) = (x_{\rm cm}(t)^2 + y_{\rm cm}(t)^2 
+ z_{\rm cm}(t)^2)^{1/2} $
denote the coordinate of the nanoparticle center of mass (CM)
at a time $t_0$ and at a later time $t$, respectively. The MSD is
 related to the  diffusion coefficient $D$ by
\begin{equation}
 D = \lim_{t \rightarrow \infty} \frac{\langle \Delta \vec r_{\rm cm}(t)^2 \rangle}{6t}\;.
\end{equation}

The structure of the fluid was analyzed using the radial distribution 
function (RDF) $g(r_{ij})$, and the pressure was evaluated with the 
virial expansion. 
In order to check if the Janus system shows density anomaly we evaluate the 
temperature of maximum density (TMD). Using thermodynamical relations, the
TMD can be characterized by the minimum of the pressure versus
temperature along isochores,
 \begin{equation}
  \left(\frac{\partial p}{\partial T}\right)_{\rho} = 0\;.
  \label{TMD}
 \end{equation}
\noindent The fluid and micellar region in the $p\times T$ phase diagram 
were defined analyzing the structure with $g(r_{ij})$,
snapshots and the diffusion coefficient $D$.

\section{Results and Discussion}
\label{Results}

  \begin{figure}[ht]
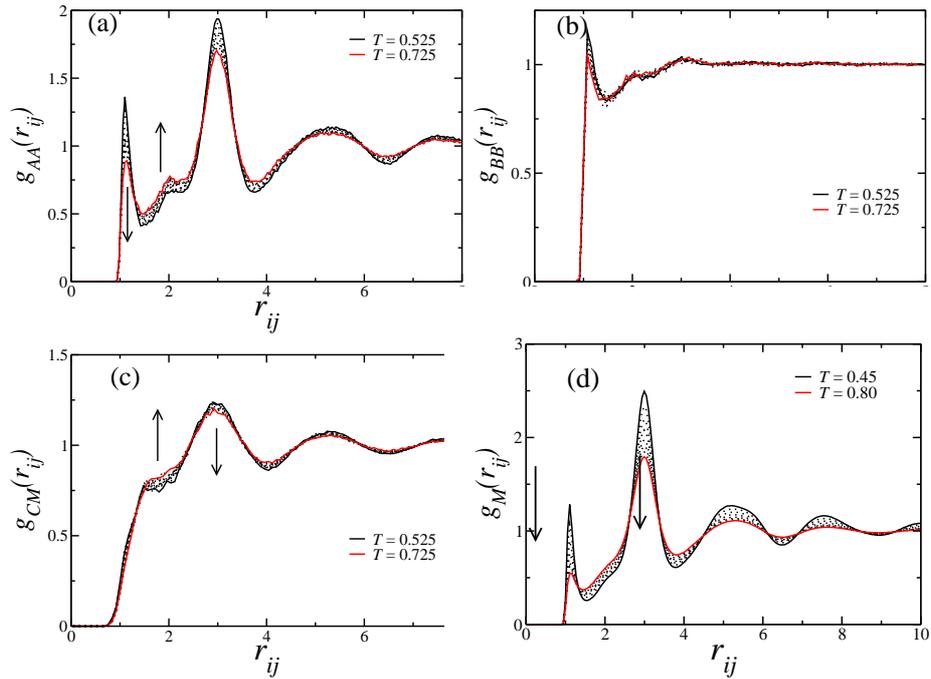

  \begin{center}
  \includegraphics[width=6cm]{fig2a.eps}
  \includegraphics[width=6cm]{fig2b.eps}
  \includegraphics[width=6cm]{fig2c.eps}
  \includegraphics[width=6cm]{fig2d.eps}
  \end{center}
  \caption{ Radial distribution function between (a) A-A monomers, (b)
  B-B monomers and (c) nanoparticles center of mass at density $\rho = 0.20$
  and temperature ranging from $T=0.525$ to $T=0.725$. (d) RDF
  for the monomoric case at density $\rho = 0.07$
  and temperature ranging from $T=0.45$ to $T=0.80$.}
  \label{fig2}
  \end{figure}
 
 Studies~\cite{Barraz09, Oliveira10} have shown that
 the main ingredient for a fluid present waterlike anomalies
 is not only the two length scales in the interaction, 
 but due the competition between this 
 scales. The TLS potential, equation~\ref{AlanEq},
 do not have density and diffusion anomalies in
 the monomeric case~\cite{Silva10}. In this way, we first
 analyze the radial distribution function $g(r_{ij})$
 between A-A monomers, namely $g_{AA}(r_{ij})$ and shown in figure~\ref{fig2}(a),
 B-B monomers, namely $g_{BB}(r_{ij})$ and shown in figure~\ref{fig2}(b), and between the
 center of mass of each dimer, $g_{CM}(r_{ij})$, figure~\ref{fig2}(c).
 
 The monomeric case, shown in figure~\ref{fig2}(d), 
 shows that both peaks, related to the first 
 and second shell, decreases when the temperature increases.
 This is the case where there is no competition between the scales,
 or there is no movement from one characteristic distance to another. 
 Surprisingly, for the dumbbells nanoparticles we can observe
 a clear competition between the length scales.
 
 The RDFs are shown for density $\rho = 0.20$ and temperature 
 between $T = 0.525$ and $T = 0.725$. The B particles RDF, $g_{BB}(r_{ij})$,
 have a small competition between the scales as we increase the 
 temperature at fixed density. On the other hand, $g_{AA}(r_{ij})$
 and $g_{CM}(r_{ij})$ clearly have a movement from the particles
 from one scale to another. The $g_{AA}(r_{ij})$ indicates that, as
 we increase the temperature, the particles leaves the first length scale, at $r_{ij}\approx1.0$,
 and go to a second length scale at $r_{ij}\approx2.0$. This is the
 distance where the fraction of imaginary modes from this family of TLS
 potentials has a local minimum~\cite{Oliveira10}.
 The $g_{CM}(r_{ij})$ shows the particles moving from a distance $r_{ij}\approx3.0$
 at lower temperatures to $r_{ij}\approx2.0$ at the higher temperatures.
 This change in the RDF peaks, with one increasing while the other decreases,
 is a strong indicative that the system will have waterlike anomalies.
 Also, the RDF shows that not only the A monomers are arranging to go to the 
 preferred distance, $r_{ij}\approx2.0$, but the center of mass also moves to the distance
 $r_{ij}\approx3.0$. This separation is equal to the second length scale plus the OLS monomer diameter,
 showing how the competition is generated by the dumbbell anisotropy.

  \begin{figure}[ht]
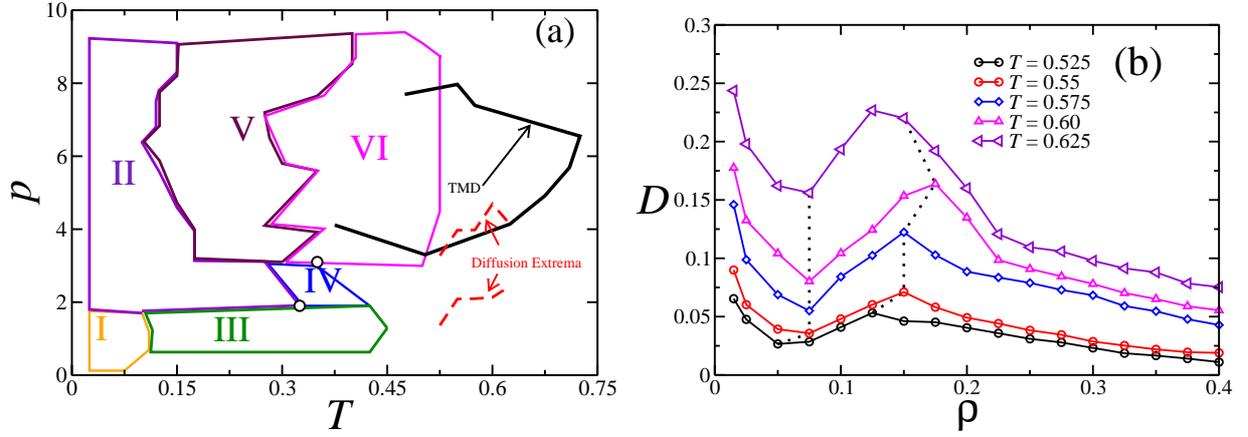

  \begin{center}
  \includegraphics[width=8cm]{fig3a.eps}
  \includegraphics[width=8cm]{fig3b.eps}
  \end{center}
  \caption{(a) Qualitative $p\times T$ phase diagram with the anomalies regions and the self-assembly structures and
  thermodynamic phases. Region I is a amorphous solid phase, with dimeric, trimeric and tetrameric
  micelles. Region II is a amorphous solid phase with spherical and wormlike micelles. Regions III
  correspond to a gas-liquid coexistence region, while region IV is a solid-fluid coexistence region.
  In the region V the system is structured in a lamellar solid micelles, and in region VI the system
  is structured in lamellar micelles with crystal-liquid characteristics. The two black circles
  corresponds to triple points. For simplicity, the isochores are not shown. (b) $D\times\rho$ curve,
  showing the diffusion maxima and minima.}
  \label{fig3}
  \end{figure}

 Therefore, as a consequence from the competition between the two length scales
 the Janus dumbbell have density and diffusion anomalies.
 In figure~\ref{fig3}(a) we show the $p\times T$ phase diagram with
 the anomalies regions. The TMD is represented by the solid black line, 
 the diffusion maxima and minima are represented by the red dashed line.
 The anomalous region does not have a waterlike hierarchy, and even an
 silicalike hierarchy. The diffusion anomalous region is at lower
 pressures (or lower densities) than the density anomalous region.
 Fomin and coworkers have shown that distinct hierarchy can be observed 
 accordingly with the parameters
 used for core softened fluids~\cite{Fomin11, Fomin13}.
 In this way, is possible to obtain the desired hierarchy
 with the potential parametrization.
 
 Figure~\ref{fig3}(b) is the 
 diffusion dependence with the system density, showing the 
 anomalous increment in $D$ as we increase the fluid density.
 This results indicates that non-anomalous material can be combined
 to create anomalous nanoparticles. 
 The small attractive well in the LJ potential is located in the same place that the
 shoulder in the TLS potential, see figure~\ref{fig2}.
 Essentially, the OLS monomers affects 
 the competition between the length scales, 
 taking the Janus dimer to have waterlike features.

  \begin{figure}[ht]
  \begin{center}
  \includegraphics[width=4cm]{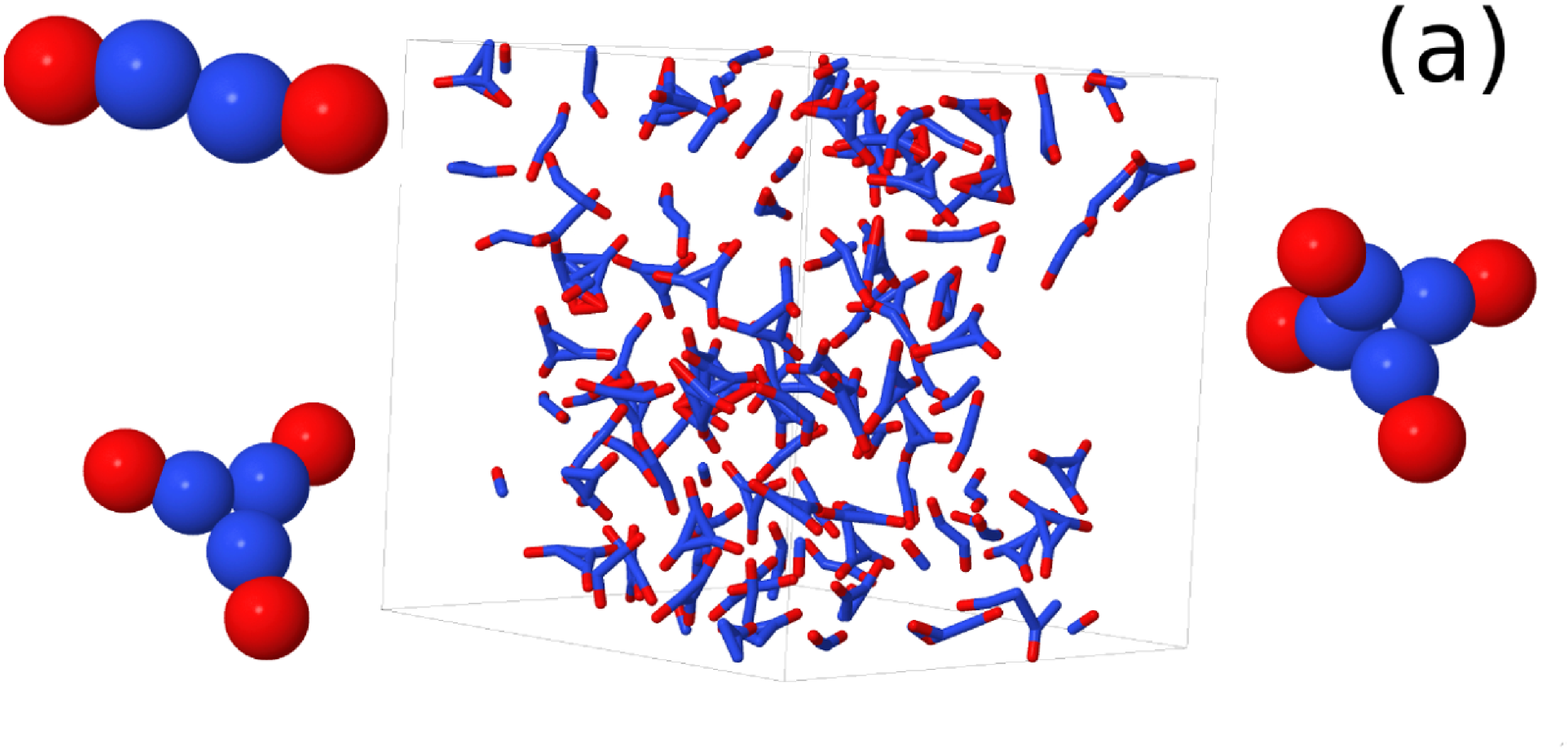}
  \includegraphics[width=4cm]{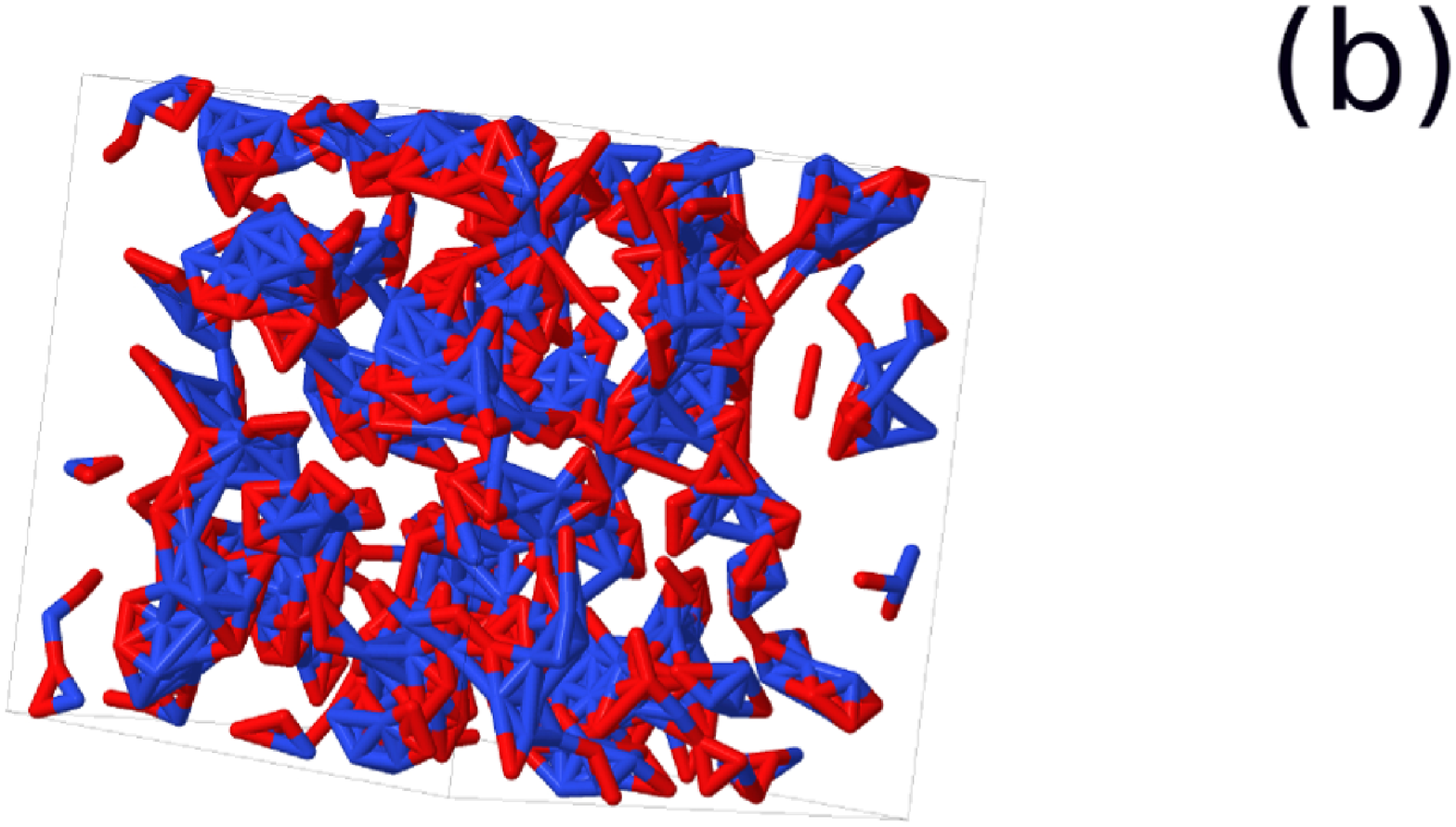}
  \includegraphics[width=4cm]{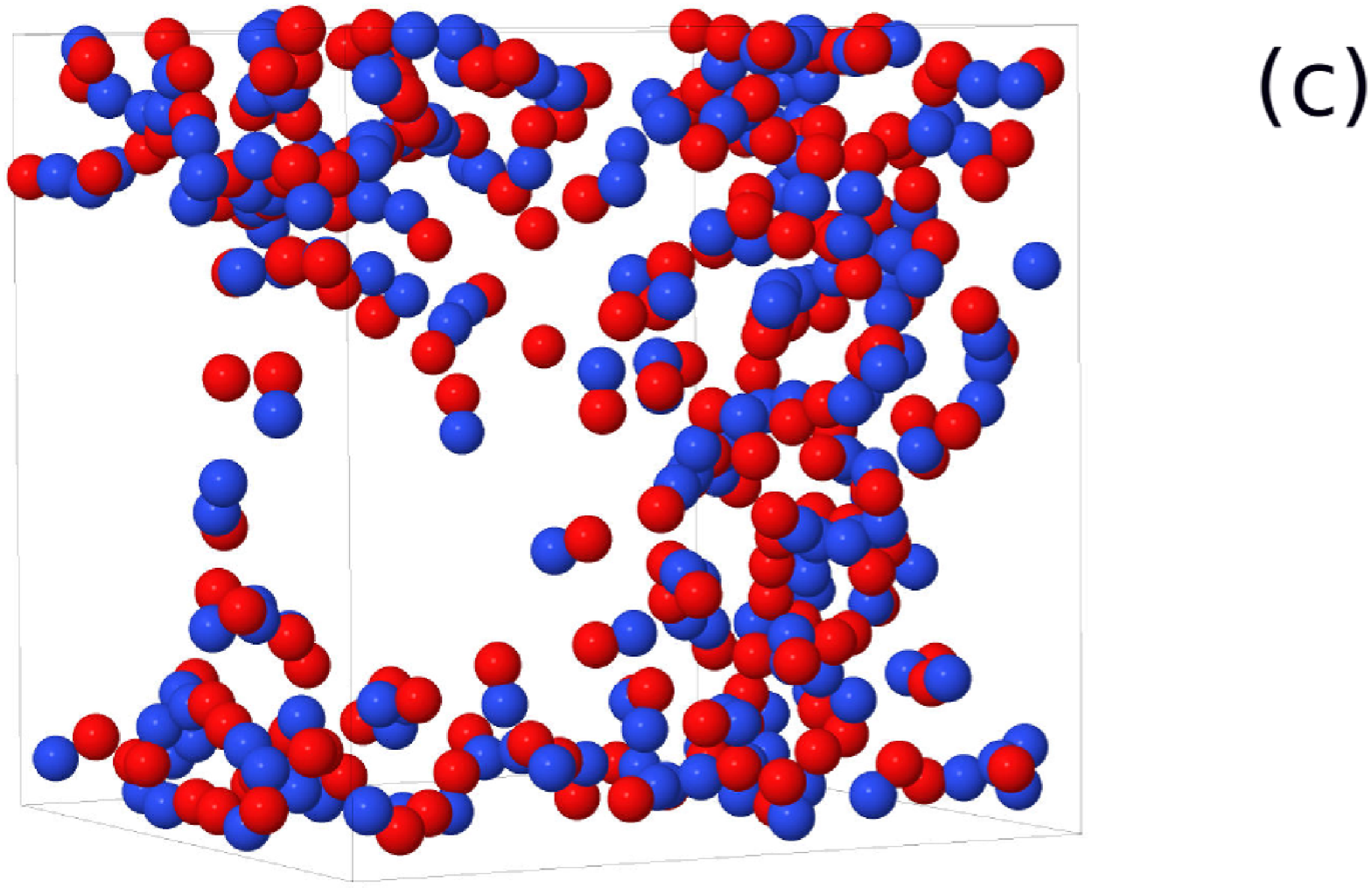}
  \includegraphics[width=4cm]{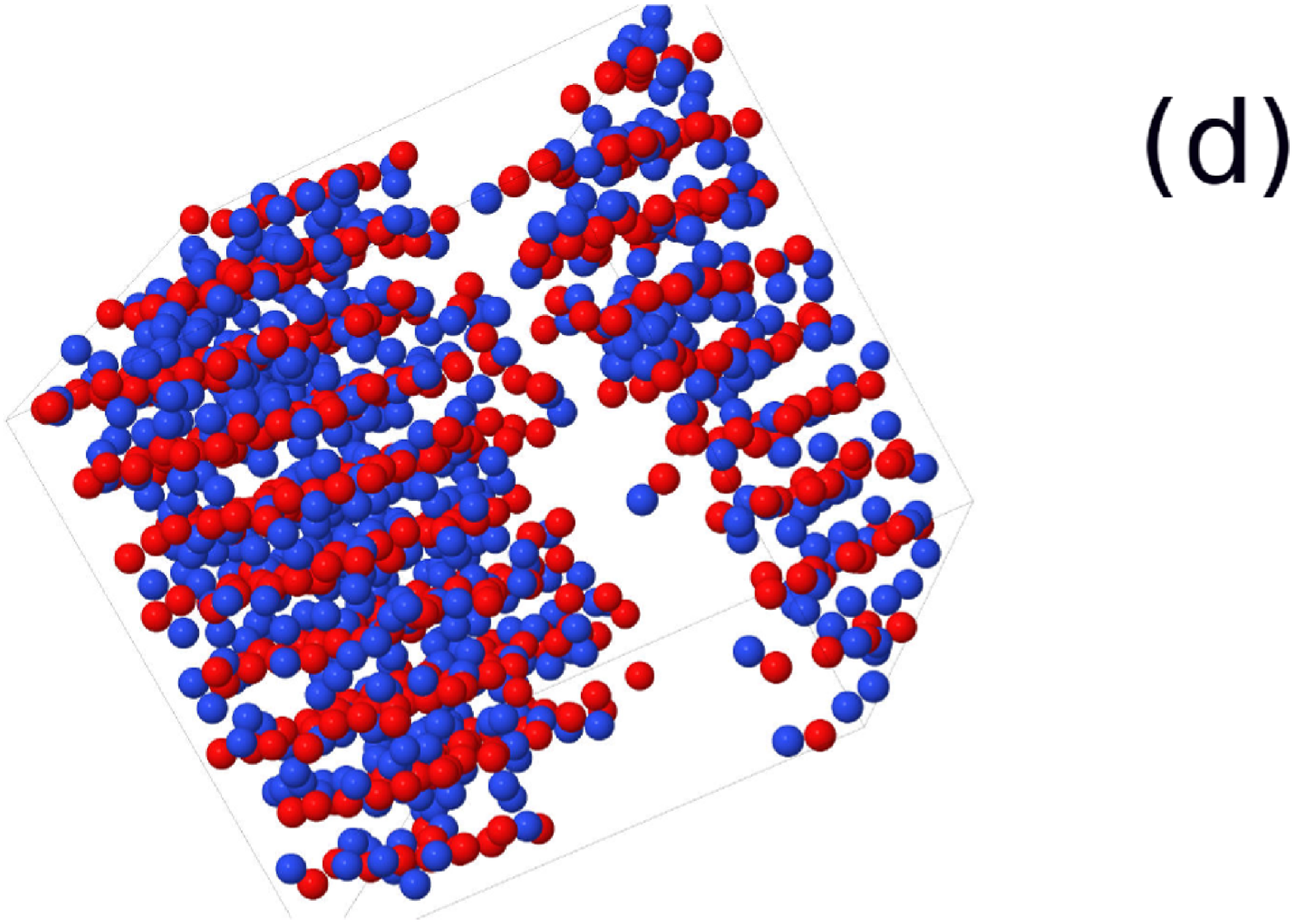}
  \includegraphics[width=4cm]{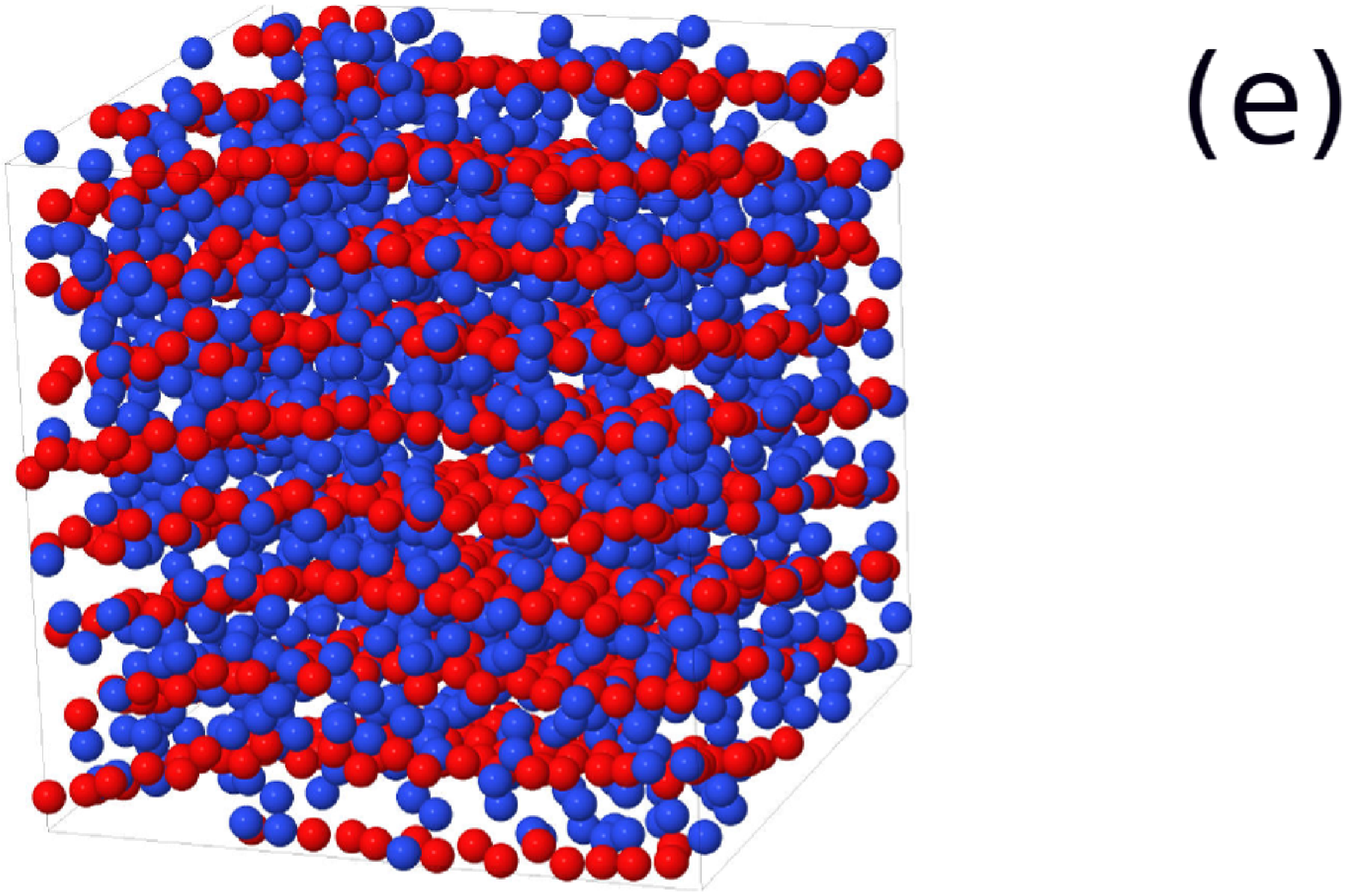}
  \includegraphics[width=4cm]{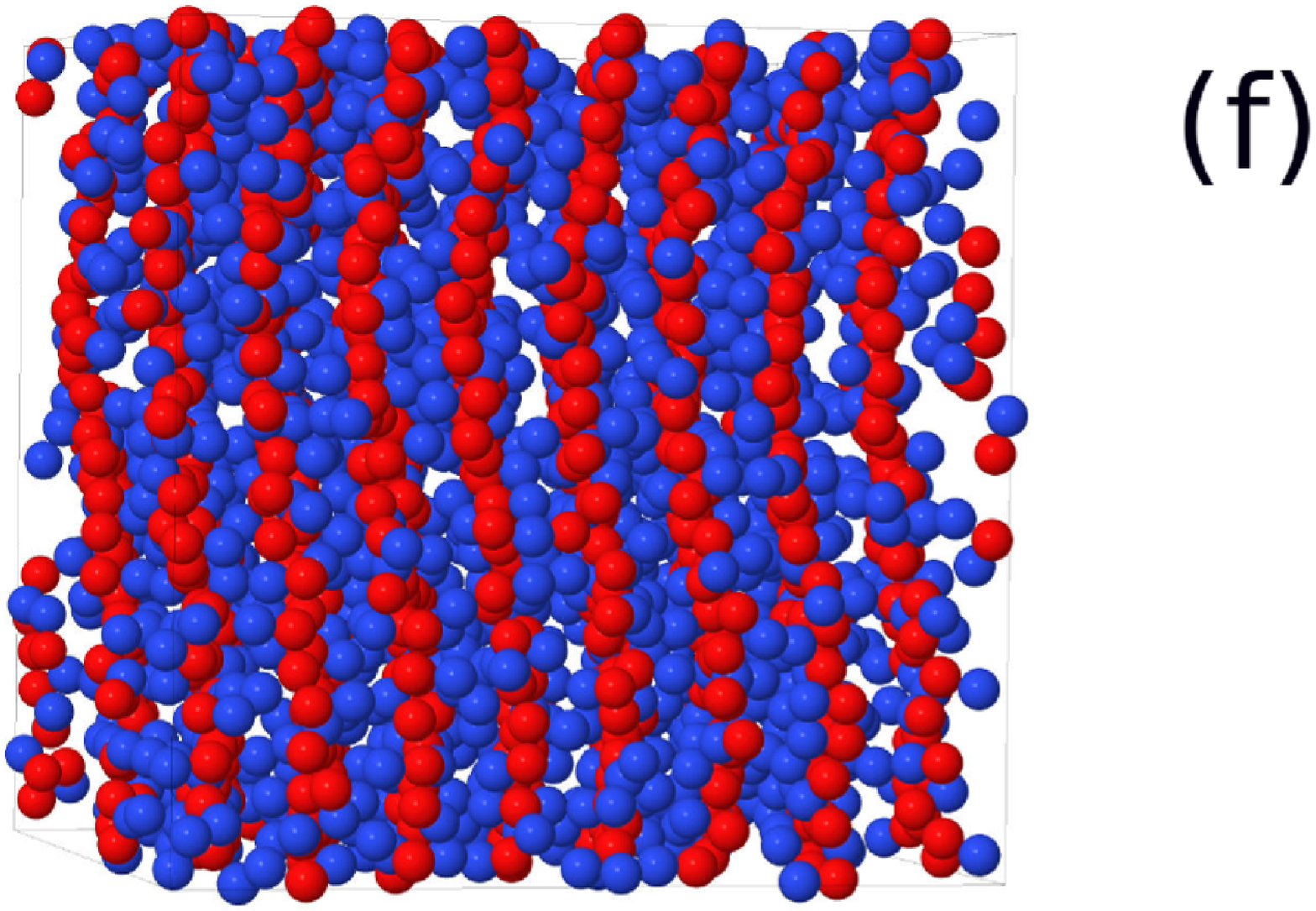}
  \end{center}
  \caption{Snapshots of the observed structures and phases. (a) is in region I
  in the phase diagram, at $T = 0.025$ and $\rho = 0.075$, (b) is in
  the region II at $T = 0.025$ and $\rho = 0.25$, (c) is in region III
  at $T = 0.10$ and $\rho = 0.05$, (d) is in region IV at 
  $T = 0.40$ and $\rho = 0.15$, (e) is in region V at $T = 0.25$ and $\rho = 0.30$
  and (f) is in region VI at $T = 0.50$ and $\rho = 0.30$}
  \label{fig4}
  \end{figure}
 
 Among the anomalous properties, the system exhibit several self-assembled
 structures. A region was defined as solid or fluid
 based in the nanoparticles structure, analyzed with the RDF and the snapshots, 
 and based in the inclination of the MSD curve. 
 With this, we should address that the phase diagram is qualitative,
 based on direct observation of the various assembled structures.
 At low temperatures and low densities,
 the orange region I in the $p\times T$ phase diagram, the nanoparticles are structured in 
 an amorphous solid state, with the dimers assembled in dimeric, trimeric
 and tetrameric clusters. The structures observed are shown in figure~\ref{fig4}(a).
 Increasing the density, more nanoparticles will agregate in the same cluster. As consequence,
 spherical and elongated (wormlike) micelles were observed, as we shown in figure~\ref{fig4}(b).
  Due the low temperature, the system remains in a amorphous solid state. 
  In order to make the snapshot more clear, only the bonds inside each micelle are shown in this case.
 At intermediate temperatures and small densities, the system exhibits cavitation and a 
 gas-liquid coexistence, region III in figure~\ref{fig4}(a), 
 as we shown in figure~\ref{fig4}(c), while at the region
 IV there is a solid-fluid coexistence, figure~\ref{fig4}(d). The holes in region III
 are similar to the observed in a model for alcohols~\cite{Munao15} that uses a
 dimer related to our model.
 The solid in region IV corresponds to a lamellar phase, where the system is structured
 in planes, with a fluid hole. 
 The lamellar structure was also observed in the regions V and
 VI, as discussed bellow in more details. The point where the
 regions II, III and IV connect, the lower circle in figure~\ref{fig3}, 
 can be understand as a triple point, since is the connection of 3 distinct
 regions. In fact, the point is crossed by several isochores, witch reinforces
 the existence of this triple point. 
 For simplicity and in order to make the $p\times T$ phase diagram more
 clear we do not shown the isochores in figure~\ref{fig3}.

  \begin{figure}[ht]
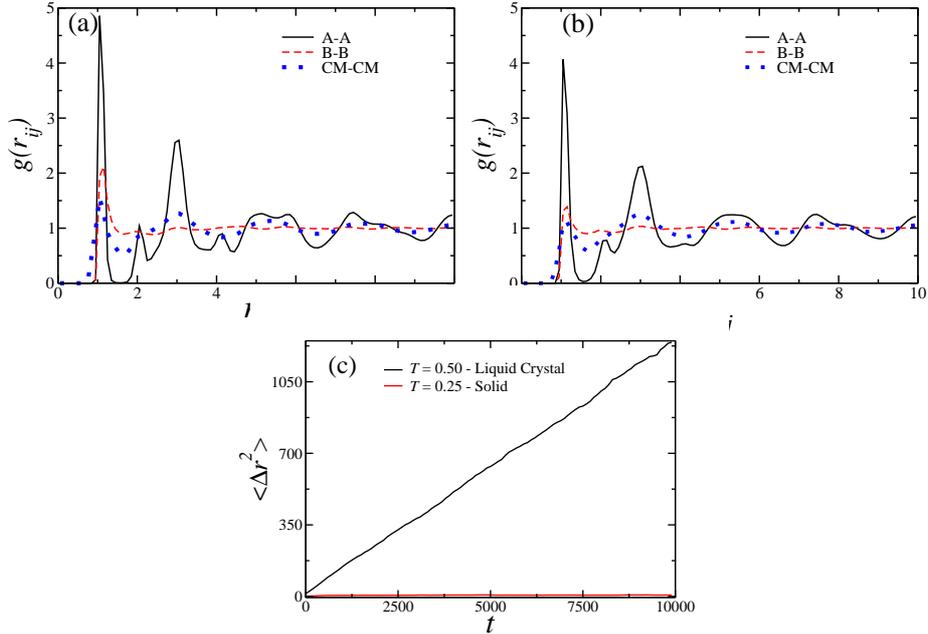

  \begin{center}
  \includegraphics[width=6cm]{fig5a.eps}
  \includegraphics[width=6cm]{fig5b.eps}
  \includegraphics[width=6cm]{fig5c.eps}
    \end{center}
  \caption{ (a) RDF for the region (V) at  $T = 0.25$ and $\rho = 0.30$ and (b) 
  for the region (VI) at $T = 0.50$ and $\rho = 0.30$. The solid black line
  is the RDF between A monomers, the dashed blue line between B monomers
  and the dotted red line between the nanoparticles center of mass.(c) Nanoparticle center of mass
  MSD at $\rho = 0.30$ for $T = 0.25$, solid red line, and $T = 0.50$, solid black line.}
  \label{fig5}
  \end{figure}
 
 The region V delimit a lamellae region. The fluid is well structured
 in planes, as the figure~\ref{fig4}(e) shows. The same kind of structure was observed
 in the region VI, figure~\ref{fig4}(f). The A monomers are fixed related to each other, 
 as the RDFs from figure~\ref{fig5}(a) and figure~\ref{fig5}(b) indicates, 
 while the B monomers do not have a well defined structure.
 In a first moment, the RDFs and the snapshots could indicate that for the same isochore,
 $\rho = 0.30$, in the region V ($T=0.25$) and
 VI ($T=0.50$) the system have the same behavior. However, analyzing the nanoparticle center of mass 
 MSD, shown in  figure~\ref{fig5}(c), we can see that at $T=0.25$ the system do not moves,
 as in a solid, while for $T=0.50$ the fluid have a large diffusion. In this way, 
 in the region VI the fluid is structured in lamellar micelles, and moves in the direction
 of this planes. This planar liquid-crystal phase was already observed for dumbbells particles
 with potentials similar to the equation~\ref{AlanEq}, but without the attractive well~\cite{Oliveira10b}.
 Also, liquid-crystal was used in recent experiments for Janus dumbbells~\cite{Jeong15}.
 In the experimental system one monomer have a polymer compartment and a larger boundary. 
 In fact, we can imagine this polymer compartment monomer as a TLS monomer, similar to 
 our model.  Also, a second triple point crossed by several isochores was found in the 
 limits between the region IV,  region VI and the region where the system is fluid.

 The variety of self-assembled structures are not surprising. 
 The micelles are present in hydrophobic-hydrophilic molecular systems,
 as surfactants and Janus particles. In our case, the competition between 
 the TLS and OLS potential creates the micelles. As usual for competing interaction
 models, the transition between the fluid to structured phases are first-order.

\section{Conclusion}
\label{Conclu}

The pressure versus temperature phase diagram of a Janus dumbbells
model was studied. We analyze the effect of an attractive two length
scales potential in the phase behavior. We have shown that, despite
the monomeric system do not exhibit waterlike anomalies, the 
dimeric system with a OLS monomer and a TLS monomer will exhibit
competition between the characteristic distances and, consequently,
density and diffusion anomalies. The model shows a rich
variety of micelles, similar to observed in 
hydrophobic-hydrophilic Janus dumbbells nanoparticles.
Regions of gas-liquid and solid-fluid coexistence were
found, both related with triple points. Also, a
liquid-crystal phase was obtained, and the TMD line
goes from the fluid region to the liquid-crystal region.

Our results indicates that is possible to create dimeric
particles with anomalous properties using non-anomalous
monomers, as colloids that have two characteristic
scales for the interaction, and still have the self-assembly
with different micellar conformations. Further investigation,
including asymmetric monomers and distinct 
OLS/TLS potentials effects on anomalies and in the hierarchy
are currently in progress.

\section{Acknowledgments}

We thank the Brazilian agency CNPq for the financial support.

\section*{References}

\end{document}